\documentstyle[12pt]{article}
\catcode`@=11
\def\address{\m@th\@ifnextchar[\@address{\@address[]}}

\def\@address[#1]#2{
\expandafter\def\expandafter\@addressname\expandafter
{\@addressname{
  \adr{#1}\ \parbox[t]{5in}{
     \ignorespaces #2}\par }}}
\def\@addressname{}
\def\adr#1{{\normalsize\unskip$^{#1}$}}
\def\@maketitle{%
\def\and{{\rm and}}
  \newpage
  \null
  {\centering
  \let \footnote \thanks
    {\Large\bf   \@title \par}%
    \vskip 1.5em%
      \lineskip .5em%
    {\bf\normalsize   \@author\par}
      \vspace{1em} 
    {\small \@addressname}
    
  }%
  \par
  \vskip 1.5em}
\def\section{\@startsection {section}{1}{\z@}{-3.5ex plus-1ex minus
    -.2ex}{1.5ex plus.2ex}{\reset@font\large\bf}}
\def\subsection{\@startsection{subsection}{2}{\z@}{-3.25ex plus-1ex
    minus-.2ex}{1.5ex plus.2ex}{\reset@font\normalsize\bf}}
\def\subsubsection{\@startsection
     {paragraph}{4}{\z@}{3.25ex plus1ex minus.2ex}{-1em}{\reset@font
     \normalsize\bf}}
\catcode`@=12

\title{Generalized Poisson structures
\footnote{Talk at XXI Int. Coll. on Group Theoretical Methods in 
Physics (Goslar, July 1996). To appear in the proceedings.} 
}
\author{J.A. de Azc\'arraga \adr{1} \footnote{St. John's College Overseas 
Visiting Scholar.}\ \footnotemark[4] , A. M. Perelomov \adr{2} \footnote
{On leave of absence from Inst. for Theor. and Exper. Phys., 
117259 Moscow, Russia.} \hskip 0pt\ \and\ 
J.C. P\'erez Bueno \adr{1} \footnote{
On sabbatical (J.A.) leave and on leave of absence (J.C.P.B.)
from Departamento de F\'{\i}sica Te\'orica and IFIC
(Centro Mixto Univ. de Valencia-CSIC), E--46100 Burjassot (Valencia), Spain.}
}
\address[1]{Department of Applied Mathematics and Theoretical Physics, 
Silver St., Cambridge CB3 9EW, UK}
\address[2]{Departamento de F\'{\i}sica Te\'orica, Univ. Zaragoza, 50009 
Zaragoza, Spain}
\font\black=msbm10 scaled\magstep1
\def\ld{\mathop{\cdots}\limits}
\def\R{{\hbox{{\black R}}}} 
\def\g{{\cal G}}
\def\be{\begin{equation}}
\def\ee{\end{equation}}
\def\ie{{\it i.e.}}
\def\eq#1{(\ref{#1})}
\begin{document}
\maketitle
\begin{abstract}
New generalized Poisson structures are introduced by using 
skew-symmetric contravariant tensors of even order. 
The corresponding `Jacobi identities' are given by the vanishing of the 
Schouten-Nijenhuis bracket.
As an example, we provide the linear generalized Poisson structures which
can be constructed on the dual spaces of simple Lie algebras.
\end{abstract}

\section{Standard Poisson structures}

We report here on the generalized Poisson structures (GPS) introduced in 
\cite{APPB}. They are different from these of 
Nambu \cite{Na} and Takhtajan \cite{Ta}. 

Let us start by recalling how the standard Poisson structures are introduced.
Let $M$ be a manifold and ${\cal F}(M)$ be the associative algebra 
of smooth functions on $M$.

\medskip
\noindent
{\bf Definition 1.1 (PB)\quad}
A {\it Poisson bracket} $\{ \cdot ,\cdot \}$ on ${\cal F}(M)$ 
is a bilinear mapping ${\cal F}(M)\times{\cal F}(M)\to {\cal F}(M)$,
$(f_1,f_2)\mapsto \{f_1,f_2\}$ such that $\{\cdot,\cdot\}$ is
(a) skew-symmetric, (b) satisfies the Leibniz rule (derivation property)
and (c) the Jacobi identity (JI).

$M$ is then called a {\it Poisson manifold}.
Because of (a), (c) the space ${\cal F}(M)$ endowed with a PB
$\{ \cdot ,\cdot \}$ becomes an (infinite-dimensional) Lie algebra.
Let $x^{j}$ be local coordinates on $U\subset M$ and consider a PB
of the form
\be
\{f(x), g(x)\} = \omega ^{jk}(x)\partial _{j}f\partial _{k}g\quad,\quad 
\partial _{j} = {\partial \over {\partial x^{j}}}\quad,\quad
j,k=1,\ldots, n={\hbox{dim}}M 
\label{cua}
\ee
Then, $\omega^{ij}(x)$ defines a 
PB if
$\omega ^{ij}(x) = -\omega ^{ji}(x)$
[(a)] and [(c)]
\be
\omega ^{jk}\partial _{k}\omega ^{lm} + \omega ^{lk}\partial _{k}\omega 
^{mj} + \omega ^{mk}\partial _{k}\omega ^{jl} = 0\;.
\label{sex}
\ee
The requirements (a) and (b) above indicate that the PB may be given in terms of 
a skew-symmetric bivector field ({\it Poisson bivector}) 
$\Lambda\in\wedge ^2(M)$ which is uniquely defined. 
Locally,
\be
\Lambda = {1\over 2} 
\omega ^{jk}\partial _j\wedge \partial _k\quad.\label{nueva}
\ee
Condition \eq{sex} is taken into account \cite{Lich} by requiring 
the vanishing of the Schouten-Nijenhuis bracket (SNB) \cite{Sc} of $\Lambda$ 
with itself, $[\Lambda, \Lambda]=0$.
Then, $\Lambda$ 
defines a {\it Poisson structure} (PS) on $M$
and the PB is defined by 
\be
\{f,g\}=\Lambda(df,dg)\quad,\quad f,g\in{\cal F}(M)
\quad.\label{oneone}
\ee

\section{Generalized Poisson structures}

Since (a) and (b) in Def. 1.1 are automatic for any bivector field, 
the only stringent condition that a 
$\Lambda\equiv\Lambda^{(2)}$ defining a PS must satisfy comes from the JI, 
eq. \eq{sex} or $[\Lambda,\Lambda]=0$.
It is then natural to consider generalizations of the standard PS
in terms of $2p$-ary operations determined by
skew-symmetric $2p$-vector fields $\Lambda ^{(2p)}$, the case $p=1$ 
being the standard one. 
Since the SNB vanishes identically if $\Lambda'$ 
is of odd order, 
only $[\Lambda',\Lambda']= 0$ for $\Lambda'$ of even order (odd {\it degree}) 
will be non-empty. 
Having this in mind, let us introduce first the generalized Poisson bracket 
\cite{APPB}.

\medskip
\noindent
{\bf Definition 2.1 \quad} A {\it generalized Poisson bracket}
$\{\cdot ,\cdot ,\ldots ,\cdot ,\cdot \}$ (GPB) on $M$ is a $2p$-linear mapping
${\cal F}(M)\times \ld^{2p}\times{\cal F}(M)\to {\cal F}(M)$
assigning a function $\{f_1, f_2,\ldots ,f_{2p}\}$ to every set 
$f_1,\ldots ,f_{2p}\in {\cal F}(M)$
which satisfies the following conditions:

\medskip
\noindent
(a) complete skew-symmetry in $f_j$;

\medskip
\noindent
(b) Leibniz rule: $\forall f_i,g,h\in {\cal F}(M),$
\be
\{f_1,f_2,\ldots ,f_{2p-1},gh\} = g\,\{f_1,f_2,\ldots ,f_{2p-1},h\}
+\{f_1,f_2,\ldots ,f_{2p-1},g\}h\quad;
\label{fourone}
\ee

\medskip
\noindent
(c) generalized Jacobi identity: $\forall f_i\in {\cal F}(M),$
\be
{1\over (2p-1)!}{1\over (2p)!} 
{\hbox{Alt}}\,\{f_1,f_2,\ldots ,f_{2p-1},\{f_{2p},\ldots ,f_{4p-1}\}\} = 0\quad.
\label{fourtwo}
\ee

Conditions (a) and (b) imply that our GPB may be given in terms of a
skew-symmetric multiderivative \ie, by a completely skew-symmetric
$2p$-vector field $\Lambda ^{(2p)}\in \wedge^{2p}(M)$.
Condition \eq{fourtwo} (to be compared with that in \cite{Na,Ta})
will be called the {\it generalized Jacobi identity} (GJI) 
and implies that $\Lambda ^{(2p)}$ must satisfy 
$[\Lambda ^{(2p)}, \Lambda ^{(2p)}]=0$.
In local coordinates the GPB has the form
\be
\{f_1(x), f_2(x),\ldots ,f_{2p}(x)\}=\omega _{j_1j_2\ldots j_{2p}}(x)
\partial^{j_1}f_1\,\partial^{j_2}f_2\,\ldots \,\partial^{j_{2p}}f_{2p}\;\quad.
\label{fourthree}
\ee
where 
$\omega_{j_1j_2\ldots j_{2p}}$ 
are the coordinates of a completely skew-symmetric tensor which,
as a result of \eq{fourtwo}, satisfies ({\it cf.} eq. \eq{sex})
\be
{\hbox{Alt}}\,(\omega _{j_1j_2\ldots j_{2p-1}k}\,\partial ^k\,
\omega_{j_{2p}\ldots j_{4p-1}}) = 0\quad.
\label{fourfour}
\ee

\medskip
\noindent
{\bf Definition 2.2\quad} \cite{APPB}
A skew-symmetric $2p$-vector field $\Lambda ^{(2p)}$, locally written as 
({\it cf.} \eq{nueva})
\be
\Lambda ^{(2p)} = {1\over {(2p)!}}\,\omega _{j_1\ldots j_{2p}}\,\partial^{j_1}
\land \ldots \land \partial ^{j_{2p}}\quad,
\label{fourfive}
\ee
defines a {\it generalized Poisson structure} (GPS) {\it iff}
$[\Lambda ^{(2p)},\Lambda ^{(2p)}]=0$, which reproduces eq. \eq{fourfour}.

\section{Generalized dynamics}

The above GPB may be used to introduce an associated dynamics \cite{APPB}.
In it, a generalized dynamical system is described by a set of $(2p-1)$ 
`Hamiltonian' functions $H_1,H_2,\ldots , H_{2p-1}$.
The time evolution of $x_j$, $f\in{\cal F}(M)$ is then defined by
\be
\dot x_j=\{H_1,\ldots ,H_{2p-1}, x_j\}\quad,\quad
\dot f=\{H_1,\ldots ,H_{2p-1}, f\}\quad.
\label{timeevolution}
\ee
The {\it Hamiltonian vector field} associated with the $(2p-1)$ Hamiltonians 
$H_1,\ldots,H_{2p-1}$ is defined by 
$X_{H_1,\ldots,H_{2p-1}}=i_{dH_1\wedge\ldots\wedge dH_{2p-1}}\Lambda$
so that
\be
X_{H_1,\ldots,H_{2p-1}}=
\Lambda_{i_1\ldots i_{2p-1} j}
\partial^{i_1}H_1\ldots\partial^{i_{2p-1}}H_{2p-1}\partial^j\quad,
\label{hamvecf}
\ee
\be
\dot x_j=X_j=(X_{H_1,\ldots,H_{2p-1}})_j=
\Lambda_{i_1\ldots i_{2p-1} j}
\partial^{i_1}H_1\ldots\partial^{i_{2p-1}}H_{2p-1}\quad.
\label{generalsystem}
\ee
Then, $\dot f=X_{H_1,\ldots H_{2p-1}}.f\ 
(=\dot x_j{\partial f\over\partial x_j})$
is given by \eq{timeevolution}.

\medskip
\noindent
{\it Remark.}\quad
As is well known, the standard Jacobi identity among $f_1\,,\,f_2$ and $H$
is equivalent to 
${d\over dt}\{f_1,f_2\}=\{\dot f_1,f_2\}+\{f_1,\dot f_2\}$;
thus, $d/dt$ is a derivation of the PB.
The `fundamental identity' for Nambu mechanics \cite{Na} and its further
extensions \cite{Ta} also corresponds to the existence of a vector field
$D_{H_1\ldots H_{k-1}}$ which is a derivation of the Nambu bracket.
In contrast, the vector field \eq{generalsystem} above is not a derivation
of our GPB.
It should be noticed, however, that having an evolution vector field which is 
a derivation of a PB is an independent 
assumption of the associated dynamics and not a necessary one.
It is also worth mentioning that the Nambu tensor \cite{ChaTak}
$\epsilon_{i_1\ldots i_n i_{n+1}} x^{i_{n+1}}$
(in connection with the decomposability of Nambu tensors see 
\cite{ChaTak,Gauth}) may also be used here (for $n$ even) since our GJI is 
also satisfied \cite{APPB}.

\section{Linear generalized Poisson structures on the duals of simple Lie 
algebras}

Let $\g$ be the Lie algebra of a simple Lie group $G$ (compact, for simplicity).
In this case the de Rham 
cohomology ring on the group manifold $G$ is the same as the Lie algebra 
cohomology ring $H^*(\g,\R)$ for the trivial action. In its
Chevalley-Eilenberg version the Lie algebra cocycles are represented by 
bi-invariant (\ie, left and right invariant and hence closed) forms on $G$
\cite{CE}.
For instance, if using the Killing metric $k_{ij}$ 
we introduce the skew-symmetric tensor
\be
\omega_{ijk}=\omega(e_i,e_j,e_k)
:=k([e_i,e_j],e_k)=C_{ij}^{l}k_{lk}=C_{ijk}\,,\, 
(i,j,k=1,\ldots,r={\hbox{dim}}\g)
\label{VIi}
\ee
$e_i\in\g$, this defines by left translation a left-invariant (LI) three-form 
on $G$ which is also right-invariant. The bi-invariance of $\omega$ then 
reads
\be
\omega([e_l,e_i],e_j,e_k)+\omega(e_i,[e_l,e_j],e_k)+
\omega(e_i,e_j,[e_l,e_k])=0
\quad,
\label{VIii}
\ee
where the $e_i$ are now understood as LI vector fields on $G$ obtained 
by left translation from the corresponding basis of 
$\g=T_e(G)$. Eq \eq{VIii} (the Jacobi identity) thus implies 
a three-cocycle condition on $\omega$ which is in fact non-trivial as is well 
known.
If we define 
$\Lambda^{(2)}={1\over 2}{C_{ij}}^k x_k\partial^i\wedge\partial^j$ on the dual 
$\g^*$ of $\g$, we see that the condition \eq{VIii} is equivalent to  
$[\Lambda^{(2)},\Lambda^{(2)}]=0$.
Hence $\Lambda^{(2)}$ defines a (standard) linear Poisson structure.
This indicates that linear GPS on $\g^*$ may be found by looking for higher 
order cocycles.

The cohomology ring of any simple Lie algebra of rank $l$ is a free ring 
generated by the $l$ (primitive) forms on $G$ of odd order $(2m-1)$. 
These forms are associated 
with the $l$ primitive symmetric invariant tensors $k_{i_1\ldots i_m}$ 
of order $m$ which may be defined on 
$\g$ and of which the Killing tensor $k_{i_1i_2}$ is just the 
first example.
As a result, it is possible to associate a $(2m-2)$ 
skew-symmetric contravariant primitive 
tensor field, linear in $x_j$, 
to each symmetric invariant polynomial $k_{i_1\ldots i_m}$ 
of order $m$.

\medskip
\noindent
{\bf Theorem\quad}
Let $\g$ be a simple compact algebra, and let 
$k_{i_1\ldots i_m}$ be
a primitive invariant symmetric polynomial of order $m$.
Then, the tensor $\omega_{\rho l_2\ldots l_{2m-2} \sigma}$
\be
\omega_{\rho l_2\ldots l_{2m-2} \sigma}:=
\epsilon^{j_2\ldots j_{2m-2}}_{l_2\ldots l_{2m-2}}
\tilde\omega_{\rho j_2\ldots j_{2m-2} \sigma}
\;,\;
\tilde\omega_{\rho j_2\ldots j_{2m-2}\sigma}:=
k_{i_1\ldots i_{m-1}\sigma}
C^{i_1}_{\rho j_2}\ldots C^{i_{m-1}}_{j_{2m-3}j_{2m-2}}
\label{fiveone}
\ee
is completely skew-symmetric and defines a Lie algebra cocycle 
on $\g$. Furthermore
\be
\Lambda^{(2m-2)}={1\over (2m-2)!}{\omega_{l_1\ldots l_{2m-2}}}^\sigma
x_\sigma \partial^{l_1}\wedge\ldots\wedge \partial^{l_{2m-2}}
\label{fivetwo}
\ee
defines a generalized Poisson structure on $\g$. 
In particular the GPB
\be
\{x_{i_1},x_{i_2},\ldots,x_{i_{2m-2}}\}=
{\omega_{i_1\ldots i_{2m-2}}}^\sigma x_\sigma
\label{inpart}
\ee
where
${\omega_{i_1\ldots i_{2m-2}}}^\sigma$ are the `structure constants' 
defining the $(2m-1)$ cocycle and hence the generalized PS.

We refer to \cite{APPB} for the proof and for a further discussion
(see also \cite{HIGHER} for a closely related structure).
It may also be shown that different 
$\Lambda^{(2m-2)}\,,\,\Lambda^{(2m'-2)}$ tensors also commute with 
respect to the SNB and that they generate a free ring.
We shall conclude by stressing that our scheme automatically provides us with 
an infinite set of examples of linear GPS; specifically, $l$ of them for each 
simple algebra of rank $l$.
Of them, the lowest one (given by the three cocycle \eq{VIi}) 
is the standard PS associated with the Lie algebra 
$\g$, for which $\Lambda^{(2)}={1\over 2}\omega_{ij}\partial^i\wedge\partial^j$ 
where $\omega_{ij}=C_{ij}^k x_k$ and $C_{ij}^k$ are the structure constants of 
$\g$.
Whether the present GPS have applications to physical systems 
(see \cite{APPB} for a simple example), at least at the level of Nambu's 
generalization (see \cite{CHAT} and references therein), is a matter for 
further work.

\section*{Acknowledgements}
This research has been partially supported by the CICYT and the DGICYT, Spain 
(AEN96--1669, PR95--439).
J.A. and J.C.P.B. wish to thank the kind hospitality extended to them at
DAMTP and A.M.P. the financial support of the Vicerectorate for research of 
Valencia University where this work was started.
Finally, the support of St. John's College (J.A.) and
an FPI grant from the Spanish Ministry of Education and Science and the CSIC
(J.C.P.B.) are also gratefully acknowledged.



\begin{thebibliography}{99}

\bibitem{APPB}
{de Azc\'arraga, J. A.; Perelomov, A. M. and P\'erez Bueno, J. C.,
J. Phys. {\bf A29}, L151-157 (1996).
{\it The Schouten-Nijenhuis bracket, cohomology and generalized 
Poisson structures} (hep-th/9605067), to appear in J. Phys. {\bf A}}

\bibitem{Na}
{Nambu, Y.,
Phys. Rev. {\bf D7}, 2405--2412 (1973)}

\bibitem{Ta}{Takhtajan, L.,
Commun. Math. Phys. {\bf 160}, 295--315 (1994)}

\bibitem{Lich}
{Lichnerowicz, A.,
J. Diff. Geom. {\bf 12}, 253-300 (1977)} 

\bibitem{Sc}
{Schouten, J.A.,
Proc. Kon. Ned. Akad. Wet. Amsterdam {\bf 43}, 449-452 (1940);
Nijenhuis, A., 
Indag. Math. {\bf 17}, 390-403 (1955)}

\bibitem{ChaTak}
{Chatterjee, R. and Takhtajan, L., Lett. Math. Phys. 
{\bf 37}, 475-482 (1996)}

\bibitem{Gauth}
Gautheron, P.,
Lett. Math. Phys. {\bf 37}, 103-116 (1996);
Alekseevsky, D. and Guha, P,
{\it On decomposability of Nambu-Poisson tensor}, Bonn Inst. f\"ur Math.
preprint MPI/96--9 (1996)

\bibitem{CE}
{Chevalley, C. and Eilenberg, S.,
Trans. Am. Math. Soc. {\bf 63}, 85--124 (1948)} 

\bibitem{HIGHER}
{de Azc\'arraga, J. A. and P\'erez Bueno, J. C.:
{\it Higher-order simple Lie algebras} (hep-th/9605213), to appear in Commun. 
Math. Phys., and these proceedings}

\bibitem{CHAT}
{Chatterjee, R.,
Lett. Math. Phys. {\bf 36}, 117-126 (1996)} 

\end{thebibliography}
\end{document}